\documentclass[pdflatex,sn-mathphys]{sn-jnl}

\jyear{2021}%

\theoremstyle{thmstyleone}%
\theoremstyle{thmstyletwo}%

\theoremstyle{thmstylethree}%

\begin{document}

\title[KiPA22 Report]{KiPA22 Report: U-Net with Contour Regularization for Renal Structures Segmentation}


\author[1]{\fnm{Kangqing} \sur{Ye}}\email{yekangq@sjtu.edu.cn}

\author[1]{\fnm{Peng} \sur{Liu}}
\author[1]{\fnm{Xiaoyang} \sur{Zou}}

\author[1]{\fnm{Qin} \sur{Zhou}}

\author*[1]{\fnm{Guoyan} \sur{Zheng}}

\affil[1]{\orgname{Shanghai Jiao Tong University}, \state{Shanghai}, \country{China}}


\abstract{Three-dimensional (3D) integrated renal structures (IRS) segmentation is important in clinical practice. With the advancement of deep learning techniques, many powerful frameworks focusing on medical image segmentation are proposed. In this challenge, we utilized the nnU-Net framework, which is the state-of-the-art method for medical image segmentation. To reduce the outlier prediction for the tumor label, we combine contour regularization (CR) loss of the tumor label with Dice loss and cross-entropy loss to improve this phenomenon.}

\maketitle
\section{Introduction}
Three-dimensional (3D) integrated renal structures (IRS) segmentation on computed tomography angiography (CTA) images plays an important role in laparoscopic partial nephrectomy (LPN) \cite{shao2011laparoscopic}. With the advancement of deep learning techniques, various deep learning methods have proven effective for IRS segmentation \cite{he2021meta, he2020dense}. Many successful algorithms are base on the U-Net architecture \cite{heller2021state}. NnU-Net (no new U-Net) is a self-configuring method based on the U-Net architecture, which outperforms most specialized deep learning pipelines in 19 public international segmentation competition \cite{isensee2021nnu}. Although nnU-Net can produce outstanding results, there are still some outlier predictions especially for the tumor label. We introduce contour regularization (CR) loss of the tumor label into our loss function to improve this phenomenon.

\section{Method}
\subsection{Preprocessing}
The intensity value of the origin images is clipped to the [0.5, 99.5] percentiles to avoid outliers.

\subsection{Network}
\begin{figure}[H]
    \centering
    \includegraphics[width=\linewidth]{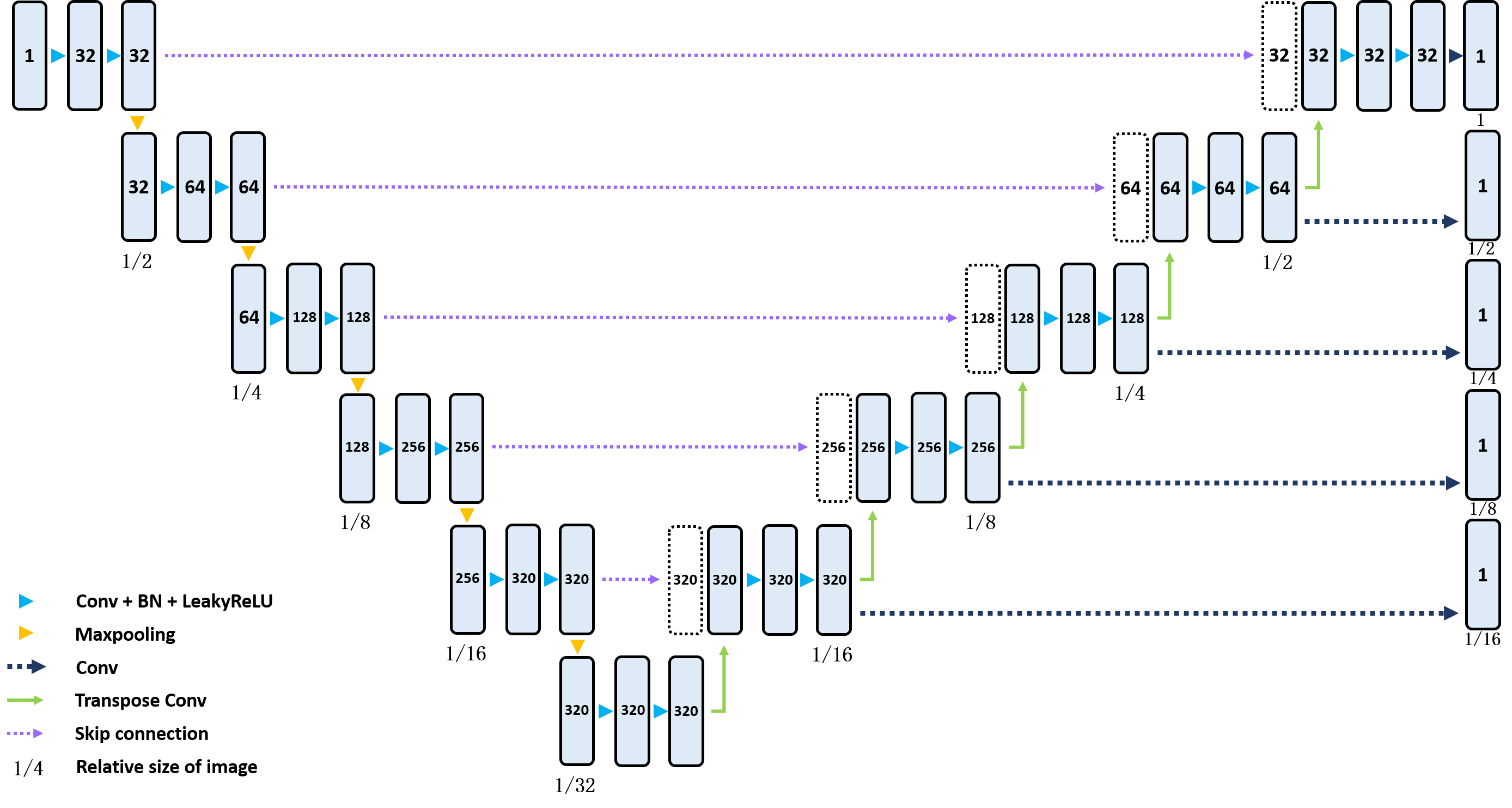}
    \caption{The 3D U-Net architecture that is used in this work. }
    \label{fig:nnU_stru}
\end{figure}

3D full resolution U-Net configuration of nnU-Net \cite{isensee2021nnu} is chosen (Fig \ref{fig:nnU_stru}). The contracting path includes $5$ downsampling steps. Deep supervision is also used.

\subsection{Loss function}
We use the summation between Dice loss, cross-entropy loss and contour regularization loss. The baseline nnU-Net method, which do not use the contour regularization loss, produces many outlier predictions of the tumor label. Contour regularization loss constrains the continuity and connectivity to reduce the outlier predictions \cite{hu2021fully}. Take bi-dimensional situation for example, let $p\in[0,1]^{H\times W}$ denote the predicted probability of each pixel to be the tumor label. The contour of prediction on position (i, j) is then defined as the largest probability variation surrounding, 

\begin{equation}
    \mathcal{C}_{d}\left(p_{i, j}\right)=\max _{\left(i_{1}, j_{1}\right),\left(i_{2}, j_{2}\right) \in \mathcal{W}_{d}(i, j)} \left\|p_{i_{1}, j_{1}}-p_{i_{2}, j_{2}}\right\|
\end{equation}
where $\mathcal{W}_{d}(i, j)=\{i \pm k, j \pm l \mid k, l \in\{0 \ldots d\}\}$ denotes the set of pixels around the point $(i,j)$ with radius $d$. The contour could be implemented as following:
\begin{equation}
    \mathcal{C}_{d}\left(p_{i, j}\right)=\max _{\substack{\left(i_{c}, j_{c}\right) \\ \in \mathcal{W}_{d}(i, j)}} p_{\left(i_{c}, j_{c}\right)}-\min _{\substack{\left(i_{c}, j_{c}\right) \\ \in \mathcal{W}_{d}(i, j)}} p_{\left(i_{c}, j_{c}\right)}=\mathcal{M}_{d}\left(p_{i, j}\right)+\mathcal{M}_{d}\left(-p_{i, j}\right)
    \label{equ:cr_1}
\end{equation}
where $\mathcal{M}_{d}$ is the max pooling operation with kernel size $2d+1$. 

Similarly, in the three-dimensional situation, the contour could be implemented as following:

\begin{equation}
    \mathcal{C}_{d}\left(p_{i, j,k}\right)=\max _{\substack{\left(i_{c}, j_{c},k_{c}\right) \\ \in \mathcal{W}_{d}(i, j,k)}} p_{\left(i_{c}, j_{c},k_{c}\right)}-\min _{\substack{\left(i_{c}, j_{c},k_{c}\right) \\ \in \mathcal{W}_{d}{(i, j,k)}}} p_{\left(i_{c}, j_{c},k_{c}\right)}=\mathcal{M}_{d}\left(p_{i, j,k}\right)+\mathcal{M}_{d}\left(-p_{i, j,k}\right)
    \label{equ:cr_2}
\end{equation}

Therefore, the contour regularization loss can be represented as following:
\begin{equation}
    \begin{aligned}
        \mathcal{L}_{CR} &=\left\|\max _{n_1,n_2 \in \mathcal{W}_{d}(n)}\left\|p_{t}^{n_1}-p_{t}^{n_2}\right\|\right\|_{2}\\
        &=\left\|\mathcal{M}_{d}\left(p_{i, j,k}\right)+\mathcal{M}_{d}\left(-p_{i, j,k}\right) \right\|_{2}
    \end{aligned}
\end{equation}
where $\left\|\cdot \right\|_{2}$ denotes $l_2$ -norm.

The whole loss function in our method is shown as below:
\begin{equation}
    \mathcal{L}_{T}  = \mathcal{L}_{Dice} + \mathcal{L}_{cross-entropy} + \alpha \mathcal{L}_{CR}
    \label{equ:cr_4}
\end{equation}
where parameter $\alpha$ adjusts the weight of contour regularization loss.
\subsection{Post-processing}
None.
\section{Dataset and Evaluation Metrics}
\subsection{Dataset}
\begin{itemize}
    \item Original images are unenhanced abdominal CT images (as .nii.gz files) from 130 patients, 70 for the training dataset, 30 for the closed testing dataset, and 30 for the opened testing dataset.
    \item The target structures are: kidney-Abnormal organ, tumor-Multi-subtype lesion, renal Artery-Very-thin structure, renal Vein-Low-significant region.
    \item In the competition phase, the model proposed in this article is trained on 70 cases in the provided training set. The provided results analysis is based on the 4-fold cross validation results.
\end{itemize}
\subsection{Evaluation Metrics}
\begin{itemize}
    \item DSC (Dice Similarity Coefficient)
    \item HD (Hausdorff Distance)
    \item AVD (Average Hausdorff Distance)
\end{itemize}
\section{Implementation Details}
\subsection{Environments and requirements}
The environments and requirements of our
method is shown in Table \ref{tab:env}.
\begin{table}[!hpt]
    \caption{Environments and requirements.}
    \label{tab:env}
    \centering
    \begin{tabular}{c|c} \hline
      OS & Ubuntu 16.04.6 LTS \\ \hline
      CPU& Intel(R) Xeon(R) Silver 4110 CPU @ 2.10GHz\\\hline
      RAM&256G\\\hline
      GPU &Nvidia RTX2080Ti\\\hline
      CUDA version & 11.1\\\hline
      Programming language&Python3.9\\\hline
      Deep learning framework & Pytorch(Torch 1.8.1, torchvision 0.9.1)\\\hline
      Specification of dependencies & nnU-Net\cite{isensee2021nnu}\\\hline
    \end{tabular}
\end{table}
\subsection{Training protocols}
Our training approach closely follows the training
method of the nnU-Net framework, which is shown in Table \ref{tab:train_pro}. We employ 4-fold cross validation strategy so that every fold has 17 or 18 validation cases. 

\begin{table}[!hpt]
    \caption{Training protocols.}
    \label{tab:train_pro}
    \centering
    \begin{tabular}{c|p{15pc}} \hline
      Data augmentation methods & Rotations, scaling, brightness, contrast, gamma correction and mirroring  \\ \hline
      Patch sampling strategy& A third of the samples in a batch contain at least one randomly chosen foreground class \cite{isensee2021nnu}.\\\hline
      Batch size&2\\\hline
      Patch size &$160\times 128\times 112$\\\hline
      Total epochs & 1000\\\hline
      Optimizer & Stochastic gradient descent\\\hline
      Initial learning rate & 0.01\\\hline
    \end{tabular}
\end{table}

\section{Results}
\subsection{Quantitative results on validation set}
Table \ref{tab:result} shows the quantitative results on the open test dataset.
\begin{table}[!hpt]
    \caption{Quantitative results on validation set}
    \label{tab:result}
    \centering
    \begin{tabular}{cccc} \toprule
      &DSC(\%)&HD(mm)&AVD(mm)\\ \midrule
      Kidney&$95.63\pm2.0$&$16.71\pm14.1$&$0.60\pm0.5$\\
      Tumor&$88.76\pm11.4$&$14.23\pm20.1$&$1.51\pm2.4$\\
      Artery&$87.33\pm4.9$&$21.44\pm24.4$&$0.46\pm0.6$\\
      Vein&$84.85\pm5.6$&$17.09\pm19.0$&$0.89\pm1.2$\\\bottomrule
    \end{tabular}
\end{table}
\subsection{Visualization result}
\begin{figure}[H]
    \centering
    \includegraphics[width=0.8\linewidth]{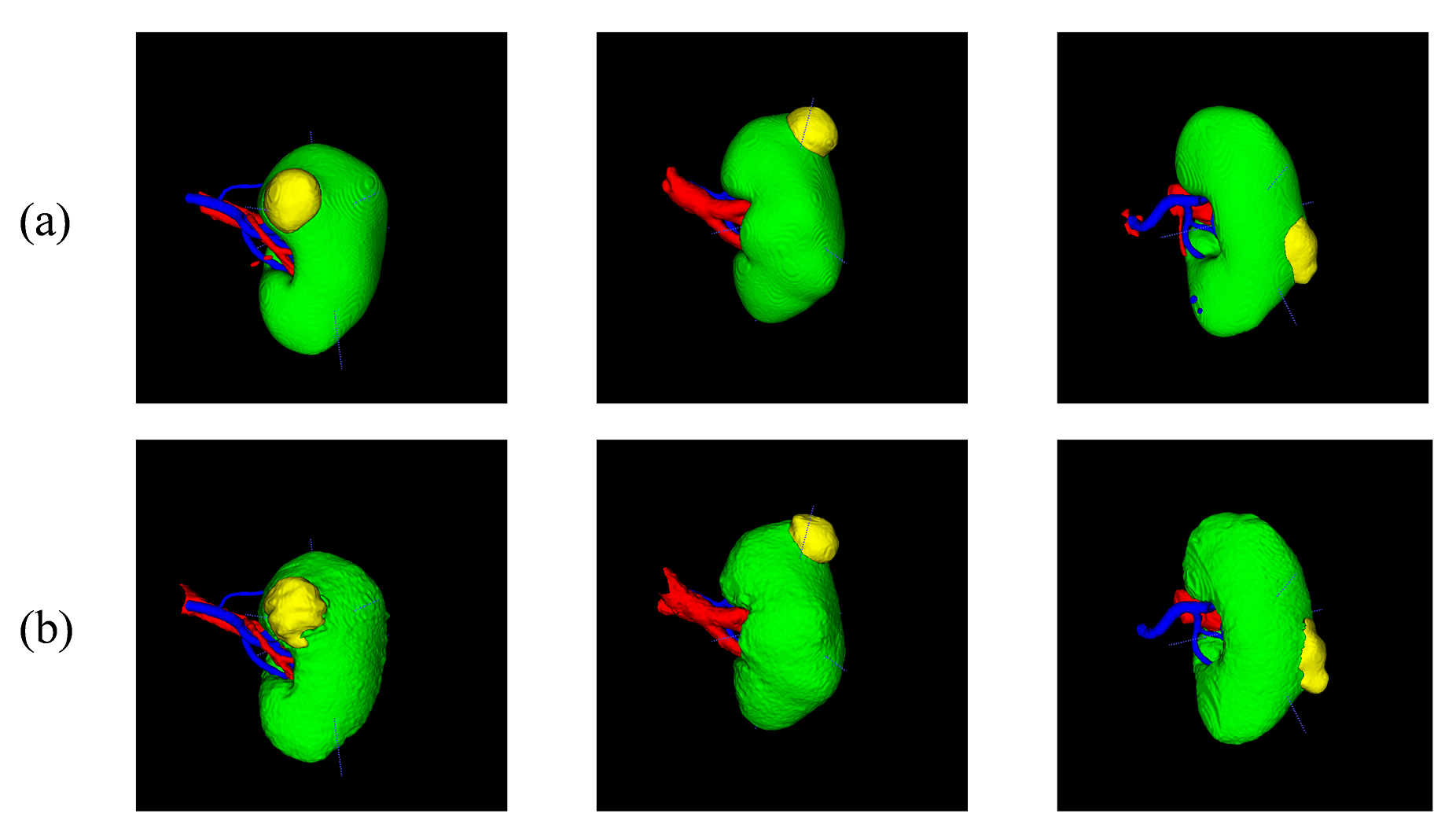}
    \caption{The visualization result of validation set in training phase. First row is the predicted result by our method, second row is the ground truth.}
    \label{fig:vis}
\end{figure}

Figure \ref{fig:vis} presents three example of validation set in training phase. 

\bibliography{sn-bibliography}


\end{document}